\documentclass[fleqn,11pt]{article}
\usepackage[utf8]{inputenc}
\usepackage[T1]{fontenc}

\usepackage{color, colortbl}
\usepackage{soul}
\usepackage{amssymb,amsmath,amsthm,enumitem}
\usepackage{graphicx}
\usepackage{csquotes}
\usepackage{float}
\usepackage{gensymb}
\usepackage[labelfont=bf, justification = justified, format = plain, figurename=Fig.]{caption}
\usepackage{cite}
\usepackage{amsmath,amssymb,amsfonts}
\usepackage{algorithmic}
\usepackage{graphicx}
\usepackage{textcomp}
\usepackage{authblk}
\usepackage[lmargin=0.75in, bmargin = 0.75in, rmargin = 0.75in, tmargin = 0.75in]{geometry}

\makeatletter
\renewcommand{\maketitle}{\bgroup\setlength{\parindent}{0pt}
\begin{flushleft}
  \textbf{\huge \@title}
\vspace{6pt}

  \@author
\end{flushleft}\egroup
}
\makeatother

\title{AI-enabled cardiac shape reconstruction from routine magnetic resonance imaging}

\author[1]{\textbf{Tanmay Mukherjee}}
\author[1]{\textbf{Neil Gautam}}
\author[2]{\textbf{Nikhil Kadivar}}
\author[3]{\textbf{Elizabeth M. Fugate}}
\author[4]{\textbf{Kyle J. Myers}}
\author[3]{\textbf{Diana Lindquist}}
\author[5]{\textbf{Pierre Croisille}}
\author[6]{\textbf{Sakthivel Sadayappan}}
\author[5]{\textbf{Patrick Clarysse}}
\author[7]{\textbf{Jacques Ohayon}}
\author[8]{\textbf{Roderic Pettigrew}}
\author[2,+]{\textbf{George Karniadakis}}
\author[1,9,10,*]{\textbf{Reza Avazmohammadi}}

\affil[1]{\small Department of Biomedical Engineering, Texas A\&M University, College Station, TX 77843, USA}
\affil[2]{School of Engineering, Brown University, Providence, RI 02912, USA}
\affil[3]{Department of Radiology, Cincinnati Children's Hospital Medical Center, Cincinnati, OH 45229, USA}
\affil[4]{Hagler Institute for Advanced Study, Texas A\&M University, College Station, TX 77843, USA}
\affil[5]{Univ Lyon, INSA-Lyon, Université Claude Bernard Lyon 1, UJM-Saint-Étienne, CNRS, Inserm, CREATIS UMR 5220, U1294, Saint-Étienne, France}
\affil[6]{Department of Cellular and Molecular Medicine, College of Medicine, University of Arizona, Tucson, AZ 85724, USA}
\affil[7]{Savoie Mont-Blanc University, Polytech Annecy-Chambéry, Le Bourget-du-Lac, France}
\affil[8]{Texas A\&M University School of Engineering Medicine, Houston, TX 77030, USA.}
\affil[9]{J. Mike Walker '66 Department of Mechanical Engineering, Texas A\&M University, College Station, TX 77843, USA}
\affil[10]{Department of Cardiovascular Sciences, Houston Methodist Academic Institute, Houston, TX 77030, USA}
\affil[+]{george\_karniadakis@brown.edu}
\affil[*]{rezaavaz@tamu.edu}

\date{}

\begin{document}


  


\maketitle


\section*{Abstract}
Computational models of cardiac structure and function are increasingly central to the development of subject-specific cardiac digital twins, enabling improved characterization of contractile dysfunction, pathological remodeling, and electrical abnormalities. A critical prerequisite for these models is the accurate reconstruction of three-dimensional (3D) cardiac anatomy from medical imaging. Multi-planar magnetic resonance imaging, particularly when combined with artificial intelligence, offers a clinically feasible alternative to conventional reconstruction techniques. In this study, we present a neural field–based reconstruction framework that recovers 3D cardiac geometries from sparse planar contour data by learning continuous shape representations.  Reconstruction performance was evaluated using complementary in-silico and in vivo datasets spanning variations in sampling density and geometric complexity. Across both datasets, reconstructed meshes closely matched reference geometries, demonstrating that the neural field approach faithfully captures cardiac planar contours. Compared with traditional local interpolation methods, the proposed framework exhibited improved geometric fidelity in anatomically challenging regions, including the left ventricular apex and basal segments, particularly under sparse sampling conditions. Collectively, these findings demonstrate that neural field–based reconstruction provides a robust and efficient pathway for multi-planar cardiac shape recovery, with particular relevance for AI-driven modeling pipelines and data-limited settings such as small-animal and time-resolved cardiac imaging.
\\[6pt]
\textbf{Keywords:} Digital twins, cardiac MRI, neural fields, shape reconstruction, shape analysis.

\newpage


\section{Introduction} \label{sec:Introduction}
Computational models of cardiac structure and function are rapidly reshaping how cardiac diseases are characterized and understood. Techniques including finite element analysis (FEA), computational fluid dynamics, and electrophysiology simulations now provide the foundational framework for building subject-specific cardiac digital twins \cite{Sel-2024, Coorey-2022, Peighambari-2026}. Indeed, computational models are increasingly recognized as a potential diagnostic indicator of contractility impairments \cite{Gu-2025}, pathological remodeling \cite{Lewalle-2025}, and conductivity disorders \cite{Qian-2025} across several disease settings. The effectiveness of these models typically relies on an appropriate simulation environment to ensure physiologically meaningful results. A critical first step toward this goal is the reconstruction of cardiac anatomy from medical imaging, herein referred to as cardiac shape reconstruction. The accuracy of this reconstruction depends strongly on imaging quality and operator capability. Historically, studies relied on modalities with high spatial resolution to inform reconstruction models \cite{Niederer-2018, Sack-2018, Mehdi-2023}. However, generating three-dimensional (3D) cardiac shapes posed practical limitations for serial acquisitions and time-resolved imaging. More recently, multi-plane or multi-view magnetic resonance imaging (MRI)–based models have emerged as effective strategies for reconstruction \cite{Mukherjee-2025SR, Xiao-2025}. When combined with super-resolution techniques and artificial intelligence (AI), MRI may provide anatomically accurate cardiac shapes in a cost-effective and clinically feasible manner

Multi-planar imaging techniques offer a comprehensive view of the in vivo anatomy and account for any deficiencies in the spatial resolution offered by a specific imaging modality. The flexibility to choose the orientation of scanning planes in MRI machines has been applied to several image processing tasks, namely super-resolution imaging \cite{Mukherjee-2025SR}, segmentation \cite{Martin-Isla-2023}, and motion tracking \cite{Meng-2022, Mukherjee-ROB-2025}. However, traditional methods for anatomical reconstruction from multi-planar MRI—such as spline-based interpolation, statistical shape modeling (SSM), and triangulation—can be laborious and time-intensive \cite{Lotjonen-2004, Chen-2020}. Consequently, AI-enabled reconstruction methods have emerged to address key limitations in (i) time consumption, (ii) reconstruction quality, and (iii) generalizability \cite{Chen-2020, Zhang-2024}. Most techniques use manually segmented in vivo heart images to train their models, while some augment this dataset with anatomically diverse shapes to improve reconstruction fidelity. In cardiac shape reconstruction, the complex motion of the heart challenges imaging accuracy and limits the availability of high-quality datasets for training learning-based AI models. As such, frameworks for generating synthetic MR images from computational models are gaining prevalence, moving beyond population-based models \cite{Mehdi-2023, Al-Khalil-2023, Wissmann-2014, Mukherjee-2024, Mukherjee-ISBI-2025}. Such in silico phantoms, thereby provide benchmarks for cardiac shapes with known deformation fields. This synthetic data strategy enables quantitative evaluation of reconstruction accuracy under controlled conditions. Moreover, this strategy is vital in small-animal imaging, where the ability to access imaging data beyond the short-axis (SAX) and long-axis (LAX) views remains largely understudied. 

In this work, we introduce NeuralMRI to address critical limitations in existing cardiac reconstruction methods through two key contributions. (i) While traditional spline-based or SSM approaches that rely on population priors or operate independently on individual contours, we implement a neural field framework that learns continuous implicit representations while leveraging global spatial context across all input planes simultaneously. This enables robust interpolation in anatomically challenging regions without requiring predefined shape spaces. (ii) Whereas deep learning (DL) methods that require extensive pre-training on large labeled datasets, our approach learns subject-specific geometry directly from sparse SAX and LAX contour data for each case, allowing generalization to diverse anatomical presentations without retraining. Additionally, we provide the first rigorous validation framework for neural field–based cardiac shape reconstruction, drawing on biomechanically realistic in-silico phantoms and in vivo murine hearts. We further extend the method to temporal reconstruction, enabling comprehensive characterization of dynamic changes in (i) dilation, (ii) contraction, and (iii) curvature across the full cardiac cycle. By transforming low-resolution, plane-specific image contours into high-resolution meshes, the proposed approach delivers automated, reproducible, and anatomically faithful cardiac shape reconstruction suitable for numerous digital-twin technologies.

\section{Related Work}
\label{sec:RelatedWork}

\subsection{Traditional learning-based methods}

Accurate 3D reconstruction of cardiac geometry from sparse imaging data remains a fundamental challenge in cardiovascular imaging. Traditional approaches have relied primarily on spline-based interpolation between planar contours or SSMs derived from population atlases~\cite{suinesiaputra2014}. Zhang et al.~\cite{zhang2016} developed an automated left ventricular reconstruction method from multiple-axis echocardiography, achieving improved accuracy through multi-view integration. However, their approach required extensive manual landmark identification, was sensitive to imaging artifacts, and struggled with regions of complex anatomical curvature. Puyol-Antón et al.~\cite{puyol2020} addressed sparse cardiac MRI reconstruction using SSMs combined with motion estimation, demonstrating improved whole-heart geometry from limited planar acquisitions. While these SSM-based methods provided computational efficiency, they fundamentally rely on predefined population-derived shape priors that may not adequately capture patient-specific anatomical variations, pathological remodeling patterns, or the fine geometric details essential for accurate biomechanical modeling~\cite{bai2018,chen2020}. Recent DL approaches have introduced data-driven alternatives to traditional reconstruction methods. Xu et al.~\cite{xu2020} proposed conditional generative adversarial networks (GANs) for cardiac MR super-resolution, while Zhou et al.~\cite{zhou2021} developed multi-view transformer-based architectures for cardiac cine MRI reconstruction. Kofler et al.~\cite{kofler2021} introduced variational networks achieving rapid functional reconstruction through learned regularization. Kong et al.~\cite{kong2021} proposed direct whole-heart mesh reconstruction using graph convolutional networks, bypassing traditional segmentation pipelines. Despite impressive reconstruction quality, these DL methods typically require extensive training on large labeled datasets with ground-truth annotations—resources that are often unavailable in clinical settings. Furthermore, these approaches may exhibit limited generalization to unseen acquisition protocols, anatomical variants, or pathological conditions that are not well-represented in the training data \cite{wang2021}.

\subsection{Neural fields for anatomical shape reconstruction}
Neural fields, or implicit neural representations, have emerged as a powerful alternative paradigm for 3D shape reconstruction, addressing several limitations of conventional approaches. Unlike methods that require large training datasets or predefined shape priors, neural fields learn continuous occupancy or signed distance functions directly from sparse input data for each case. Sawdayee et al.~\cite{sawdayee2023} introduced OReX, a neural field approach for object reconstruction from planar cross-sections that directly addresses the ill-posed problem of volumetric interpolation from sparse planar data through hierarchical sampling and iterative refinement. Marin-Castrillon et al.~\cite{marin2023} successfully applied continuous implicit neural representations to cardiac shape reconstruction from anisotropic CMR segmentations, demonstrating robustness to missing data and irregular sampling patterns. Qin et al.~\cite{qin2024} extended implicit representations to real-time cardiac cine MRI through subspace decomposition, while Wolterink et al.~\cite{wolterink2022} incorporated Fourier-feature inputs for free-breathing cardiac MRI reconstruction. Chen et al.~\cite{chen2023} demonstrated versatility across imaging modalities by applying neural signed distance fields to freehand 3D ultrasound. However, systematic evaluation of neural field methods for cardiac reconstruction from clinically relevant sparse planar sampling protocols—particularly for biomechanical modeling applications requiring known ground-truth validation—remains largely unexplored. Previous cardiac neural field studies have primarily focused on image-space reconstruction from relatively dense MRI acquisitions or have lacked rigorous quantitative validation frameworks with known geometric ground truth necessary for biomechanical simulation applications. By integrating the OReX neural field framework~\cite{sawdayee2023} with FE simulation-derived validation data, we systematically evaluate reconstruction accuracy across two clinically relevant sampling strategies: sparse protocols that mimic rapid clinical acquisitions and dense protocols that represent comprehensive research imaging. The resulting meshes exhibit geometric accuracy and element quality suitable for FE biomechanical simulation, establishing neural field reconstruction as a viable pathway for generating patient-specific cardiac geometries from clinically feasible sparse planar imaging acquisitions.


\section{Methods}
\label{sec:Methods}

\subsection{Data acquisition and preprocessing}
Two distinct datasets were utilized in training NeuralMRI: (i) an FE dataset, consisting of synthetically generated biventricular cardiac geometries, and (ii) an in vivo dataset made up of subject-specific single ventricular geometries obtained from multi-planar MRI images subjected to SRR. 

\subsubsection{FE simulations}\label{sec:fe_dataset_creation}
Synthetic cardiac shape data were generated from FE simulations of rodent-specific biventricular, i.e., left and right ventricular (LV and RV, respectively) geometries. The initial or underformed cardiac geometry was reconstructed from MRI scans and subsequently meshed using Materialize 3-Matic software with an adaptive tetrahedral mesh strategy \cite{Mehdi-2025}. The minimum element size was maintained at 0.5 mm, with inter-nodal distances at the endocardial and epicardial boundaries specified as 1 mm, measured as the length of the hypotenuse of each tetrahedral element. Subsequently, the geometry was deformed over a full cardiac cycle using a custom FE pipeline. Briefly, myofiber orientations were mapped onto the meshed geometry using a Laplace-Dirichlet rule-based algorithm.
\begin{figure}[h!]
    \centering
    \includegraphics[width=0.8\columnwidth]{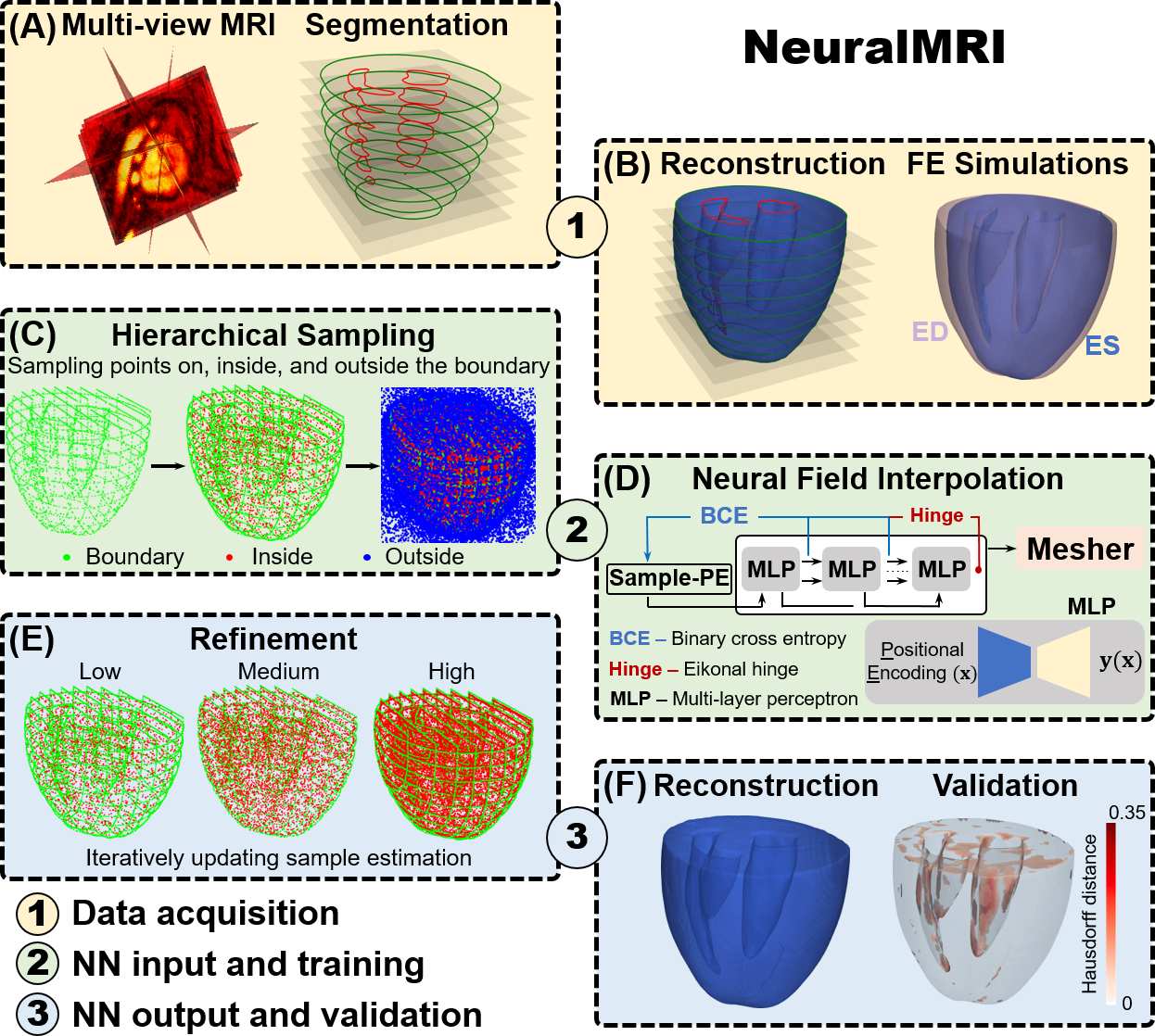}
    \caption{Flowchart describing the workflow for cardiac shape reconstruction from in-plane segmentation contours. (A) Multi-view magnetic resonance imaging (MRI) and in-plane segmentation of the datasets (mouse and human-specific) (B) Ground-truth reconstruction at end-diastole (ED) and end-systole (ES), with finite element simulations of the to augment the dataset by creating several geometries between ED and ES, (C) hierarchical sampling of points inside, outside, and on the boundary to create a geometric representation of the cardiac shape, (D) learning the input representation using a neural field (NF), (E) Refining the NF iteratively to increase the sampling of points describing the cardiac shape, (F) reconstruction through a meshing algorithm, and evaluation against ground-truth geometries.}
    \label{fig:maindiagram}
\end{figure}
The geometry was divided into six transmural layers between the endocardium and epicardium in both the LV and RV, with fiber helicity ranging from $+84^\circ$ at the endocardium to $-42^\circ$ at the epicardium to capture the physiologically realistic anisotropic myocardial architecture. Six different configurations of the fiber architecture were prescribed to generate cardiac motion with varying degrees of anisotropic bias. The myocardium was modeled as a transversely isotropic hyperelastic material, with mechanical behavior characterized by passive hyperelastic strain energy functions and active stress generation aligned with the myofiber direction \cite{Mendiola-2024}. The computational model was subjected to physiologically realistic pressure-volume loading conditions obtained through catheterization, and FE simulations were performed to compute nodal displacement fields throughout the complete cardiac cycle. Through these simulation, 500 distinct cardiac shapes were generated with several contours describing each geometry. 

\subsubsection{Multi-view MRI}\label{sec:invivo_dataset_creation}

Multi-view MRI was performed in a cohort of five six-month-old WT mice (n = 5). These mice were subjected to SAX imaging, with LAX scans also acquired via orthogonal sampling. The cardiac shape was subsequently segmented at 10 timepoints between end-diastole and end-systole (ED and ES) to capture the systolic contraction of the LV using the SRR in CMR framework presented by Mukherjee et al. \cite{Mukherjee-2025SR}. The SRR in the CMR framework combined segmented SAX and LAX images of the LV to increase the spatial resolution of the multi-view MRI scans. The spatial resolution of the MRI scans was increased from 128 $\times$ 128 $\times$ 16 voxels to 256 $\times$ 256 $\times$ 220 through scattered data interpolation. 
In addition to mouse-specific imaging, a retrospective cohort of human patients (n = 3) who underwent SAX and LAX MRI was also used to train and validate the NeuralMRI framework \cite{Duchateau-2023}. The subject-specific in vivo dataset comprising images with an average resolution of 128 $\times$ 128 $\times$ 16 voxels was segmented to isolate the LV at multiple timepoints between ED and ES. 
Using this high-resolution set of images, the LV was reconstructed through Delaunay triangulation and remeshed in 3-Matics (Materialise Inc, Leuven, Belgium). 50 distinct cardiac shapes were generated, along with several contours sampled at multiple SAX and LAX planes (Fig. \ref{fig:maindiagram}A and B).

\subsection{Dataset sampling}


From the resulting high-fidelity cardiac shape at each timepoint desribed in Sections \ref{sec:fe_dataset_creation} and \ref{sec:invivo_dataset_creation}, planar cross-sectional data were synthetically extracted along both SAX and LAX orientations. The outermost and innermost boundaries in each plane were used to define the ground-truth positions of the endo- and epicardial contour data ($\mathbf{x} \in \mathbb{R}^3$). These labeled point sets served as the primary input for the NeuralMRI pipeline, providing the geometric representation or occupancy values required for implicit shape learning (Fig. \ref{fig:maindiagram}C). The hierarchical sampling approach consisted of initializing an arbitrary point set in the local orientation of the planar contours, $\mathbf{P} \in \mathbb{R}^{3}$. A  binary occupancy distribution function was formulated to determine the location of a sampled point ($p \in P$)as it pertained to $\mathbf{x}$ as:
\begin{equation}\label{eq:occupancy}
M(p, \mathbf{x}) = \Biggl\{ \begin{matrix}
        1, \hspace{0.05in} \text{if} \hspace{0.05in} p \in \mathbf{x} \\
        0, \hspace{0.05in} \text{if} \hspace{0.05in} p \notin \mathbf{x} \\
        0.5, \hspace{0.05in} \text{if} \hspace{0.05in} p \in \partial \mathbf{x} \\
    \end{matrix}
\end{equation}
where $M$ is the occupancy distribution function and $\partial \mathbf{x}$ denotes the outermost boundary points (endo- and epicardium). Since the proposed approach leverages a combination of synthetic and in vivo image-contour pairs, repositories of point sets were generated separately for each dataset, herein referred to as the FE and in vivo datasets, respectively. 

\subsection{Neural field-based reconstruction}

The reconstruction of volumetric cardiac shapes from the multi-planar MRI contours using NeuralMRI was performed using a neural field-based approach adapted from geometric DL models for shape interpolation from sparse cross-sectional data. A multilayer perceptron, $f_\theta: \mathbb{R}^3 \rightarrow [0,1]$ parameterized by network weights ($\theta$) was trained to learn the implicit representation of a cardiac geometry by estimating an inside/outside indicator function using the corresponding positional data, i.e.,  3D Cartesian coordinates ($\mathbf{x} = (x, y, z) \in \mathbb{R}^3$). The network outputs a continuous occupancy ($y(\mathbf{x})$), indicating whether a vertex or point $P$ lies inside, on, or outside the endo- and epicardial boundaries forming the cardiac shape (Fig. \ref{fig:maindiagram}D). During training, the network was optimized to minimize the loss between the predicted occupancy values ($f_\theta(\mathbf{x})$) and the ground-truth values ($M$) from the input as:
\begin{equation}
\mathcal{L}_{\textbf{total}}(\mathbf{x}, \theta) = \mathcal{L}_{\text{BCE}} + \lambda\mathcal{L}_{\textbf{hinge}}, 
\label{eq:total_loss}
\end{equation}
where $\mathcal{L}_{\text{BCE}}$ is the binary cross-entropy (BCE), $\mathcal{L}_{\text{hinge}}$ is the Eikonal hinge loss and $\lambda$ is a hyperparameter. The two additive loss terms were defined as:

\begin{align}
\mathcal{L}_{\text{BCE}}(\mathbf{x}, \theta) = &  -\frac{1}{N}\sum_{i=1}^{N} \left[  M_i \log(f_\theta(\mathbf{x}_i)) + \notag \right. \\
&\phantom{{}=0.5} 
\left. (1-M_i)\log(1-f_\theta(\mathbf{x}_i)) \right] ,
\label{eq:bce}
\end{align}

\begin{equation}
\mathcal{L}_{\text{hinge}}(\mathbf{x}, \theta) =  \text{max} \left( \left(0,  \|\nabla f_\theta(\mathbf{x}_{N-1})\| - \alpha\right)\right),
\label{eq:Eikonal_hinge_loss}
\end{equation}
\noindent where  $N$ denotes the total number of training iterations and $\alpha$ is a hyperparameter. To address the challenge of capturing high-frequency anatomical details while maintaining smooth interpolation, an iterative estimation architecture with hierarchical input sampling was employed. 
The network was trained using a reproducible coarse-to-fine sampling strategy in which boundary regions were progressively refined across successive training stages. Training proceeded through six predefined refinement levels (levels 0–5), with each level trained sample-wise for a fixed 120 iterations, enabling systematic and repeatable refinement of fine-scale spatial features at object boundaries (Fig. \ref{fig:maindiagram}E). The number of refinements was determined through an ablation analysis, with particular emphasis on maximizing the NeuralMRI model's accuracy in reproducing the ground-truth planar contours, as explained in detail in the next section. Once the neural field was trained, the 3D isosurface representing the cardiac geometry was extracted using the marching cubes algorithm at an isovalue threshold of 0.5, yielding a triangulated isosurface $\mathcal{S} = (\mathcal{V}, \mathcal{F})$ with vertices $\mathcal{V} \subset \mathbb{R}^3$ and faces $\mathcal{F}$ suitable for evaluation (Fig. \ref{fig:maindiagram}F).

\subsection{Evaluation of spatial reconstruction}

Several well-known distance-based and accuracy metrics were used to evaluate the performance of NeuralMRI in segmenting and reconstructing the murine heart, yielding important insights into the model's overall predictive capability. The Chamfer and Hausdorff distance metrics are particularly significant for regional reconstruction and segmentation tasks. These metrics, described below, are crucial for assessing the model's ability to reconstruct local geometric features, such as curvature, smoothness, and overall similarity to the ground-truth FE simulations. To quantify the geometric robustness of the neural field model, we evaluated reconstructed surfaces using both point-set distance metrics and curvature-dependent error analysis.

\subsubsection{Geometric correspondence analysis}

Let $\mathcal{V}^{\ast}= \{v^{\ast}\}_{i=1}^{N} \subset S^{\ast}$ denote the ground-truth or reference vertex samples and $P_\theta = \{q_j\}_{j=1}^{N_\theta} \subset S$ the samples obtained from the neural field prediction. For any point $v\in \mathcal{V}^\ast$, the directed point-to-set distance ($d(v, P_{\theta}) = \min_{v\in P_\theta} \|v - x\|_2 $) can be used to compute the Chamfer distance as:
\begin{equation}
\text{Chamfer}(\mathcal{V}^{\ast}, P_{\theta}) =  \frac{1}{N_\ast}\sum_{v\in \mathcal{V}^\ast} d(v,P_\theta)^2 + \frac{1}{N_\theta}\sum_{q\in P_\theta} d(q,V^\ast)^2,
\end{equation}
where squared distances emphasize sensitivity to local deviations. We additionally report the RMS Chamfer value to maintain comparability to root-distance metrics. The Hausdorff distance can also be evaluated using ($d(p, P_{\theta}$) as: 
\begin{equation}
\text{Hausdorff}(\mathcal{V}^{\ast},P_\theta) = \max\Big( \max_{v\in \mathcal{V}^\ast} d(v,P_\theta), \;\max_{q\in P_\theta} d(q,\mathcal{V}^\ast) \Big),
\end{equation}
which captures worst-case deviations but is sensitive to isolated outliers. Therefore, to improve robustness, we also compute the percentile Hausdorff distance, where the maximum operator is replaced with the $p^{\text{th}}$ percentile, with $p \in \text{\{50,90,95\}}$), thereby emphasizing dominant shape errors rather than isolated spikes. Additionally, metrics such as, the root mean-squared error (RMSE) and mean absolute error (MAE) were also determined using the Chamfer and Hausdorff distance values to quantify overall correspondence of $\mathcal{V}^{\ast}$ and $P_\theta$.

\subsubsection{Curvature analysis}
Additionally, the geometric fidelity of the shapes was quantified using the Gaussian curvature error, which captures differences in local surface bending between the reference and predicted meshes. Let 
$K^\ast = \{K_i^\ast\}_{i=1}^{N_\ast} \quad\text{and}\quad K_\theta = \{K_j\}_{j=1}^{N_\theta},$ denote the Gaussian curvature values computed at the samples of $S^\ast$ and $S$, respectively. 
Gaussian curvature at each mesh vertex was estimated using the standard angle-deficit formulation, given by, 
    $K_i = \frac{1}{A_i}\left(2\pi - \sum_{t \in \mathcal{T}(i)} \beta_{t,i}\right)$
where $\beta_{t,i}$ are the angles incident to vertex iii in its adjacent triangles $\mathcal{T}(i)$, and $A_i$ is the associated barycentric Voronoi area. 
This discrete operator provides a consistent approximation of intrinsic curvature and is widely adopted for mesh-based geometric analysis. For any corresponding pair of surface samples ($v_i^\ast, q_i$), the pointwise curvature deviation was 
\begin{equation}
    \text{Curvature}(K^\ast, K_{\theta}) = K_i^\ast - K_i. 
\end{equation}
To extend the validation spatial reconstruction to temporal dynamics, the neural field-based reconstruction pipeline was applied independently to each time frame $t_k$ within the cardiac cycle, where $k = 1, 2, \ldots, T$ and $T$ denotes the total number of temporal phases. For each time point, the corresponding set of SAX and LAX contours was used to train or fine-tune the neural field $f_{\theta_k}$, generating a sequence of 3D meshes $\{\mathcal{M}_k\}_{k=1}^T$ representing the heart at successive phases of contraction and relaxation. Mesh correspondence across time frames was established through vertex tracking or non-rigid registration methods to enable quantitative analysis of cardiac deformation.

\section{Results}
\label{sec:Results}

\subsection{Evolution of training loss}

\begin{figure}[t!]
    \centering
    \includegraphics[width=0.8\columnwidth]{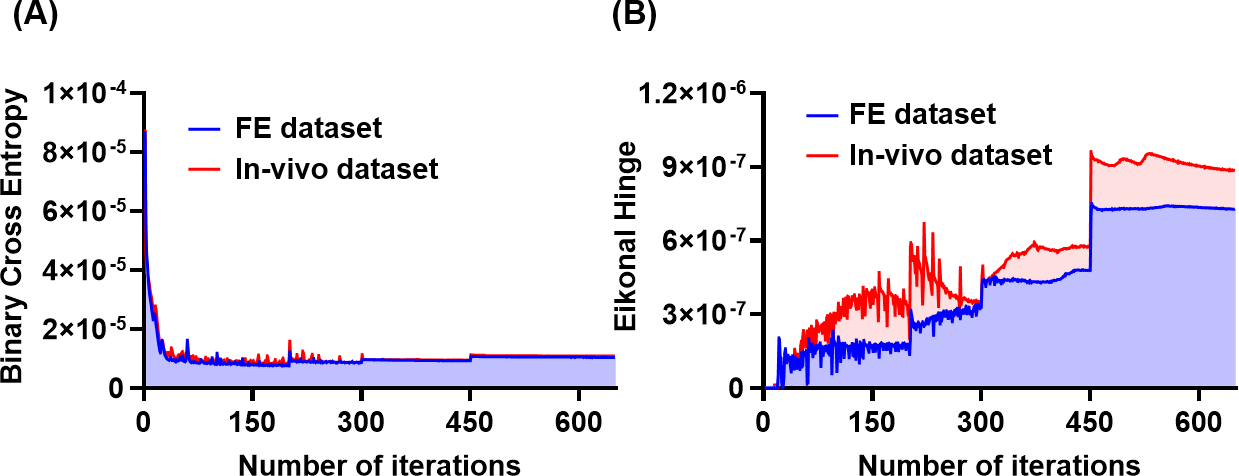}
    \caption{Evolution of training loss for the neural field for cardiac reconstruction in the finite element (biventricular mouse geometries) and in vivo (single ventricle mouse and human) datasets. Training losses are presented for (A) binary cross-entropy loss measuring the accuracy of the inside/outside indicator function, and (B) Eikonal hinge loss promoting signed distance function properties and surface smoothness. Each level of refinement (0-5) consists of 120 iterations.}
    \label{fig:loss-curves}
\end{figure}

The evolution of individual loss components during training confirmed the stability of the neural field reconstruction process. The progression of key loss terms, namely, Eikonal hinge and BCE losses, is presented as growth curves against the total number of iterations (Fig. \ref{fig:loss-curves}).
The Eikonal hinge loss, which enforces smoothness and the signed distance function property, exhibited stable convergence toward a low magnitude throughout training, with staged increases corresponding to hierarchical sampling transitions during the coarse-to-fine optimization schedule (Fig. \ref{fig:loss-curves}A). The BCE loss, which represents reconstruction accuracy, declined sharply in the early iterations before stabilizing, indicating successful learning of the inside/outside indicator function. 
\begin{table}[h!]

\renewcommand{\arraystretch}{1.2} 
\caption{Summary of metrics used to evaluate the accuracy of cardiac shape reconstruction for FE dataset (biventricular geometries) and the in vivo dataset (mouse- and human-specific single ventricle).}
\centering
\begin{tabular}{|c|c|c|}
\hline
\textbf{Metric} & \textbf{FE Dataset} & \textbf{In vivo dataset} \\
\hline
Chamfer distance & 2.7146 $\pm$ 0.0569  & 0.9443 $\pm$ 0.2028 \\
Hausdorff distance & 9.9414 $\pm$ 0.3777 & 1.3870 $\pm$ 0.2122 \\
Optimal threshold & 9.4603 $\pm$ 0.3122 & 0.6416 $\pm$ 0.1257 \\
Curvature & 9.4603 $\pm$ 0.3122 & 0.6416 $\pm$ 0.1257\\
\hline
\end{tabular}\label{tab:table_1}

\end{table}
These convergence trends demonstrate effective minimization of the reconstruction and regularization objectives, with the network learning a smooth, geometrically consistent implicit representation of cardiac anatomy.
\begin{figure}[h!]
    \centering
    \includegraphics[width=0.8\columnwidth]{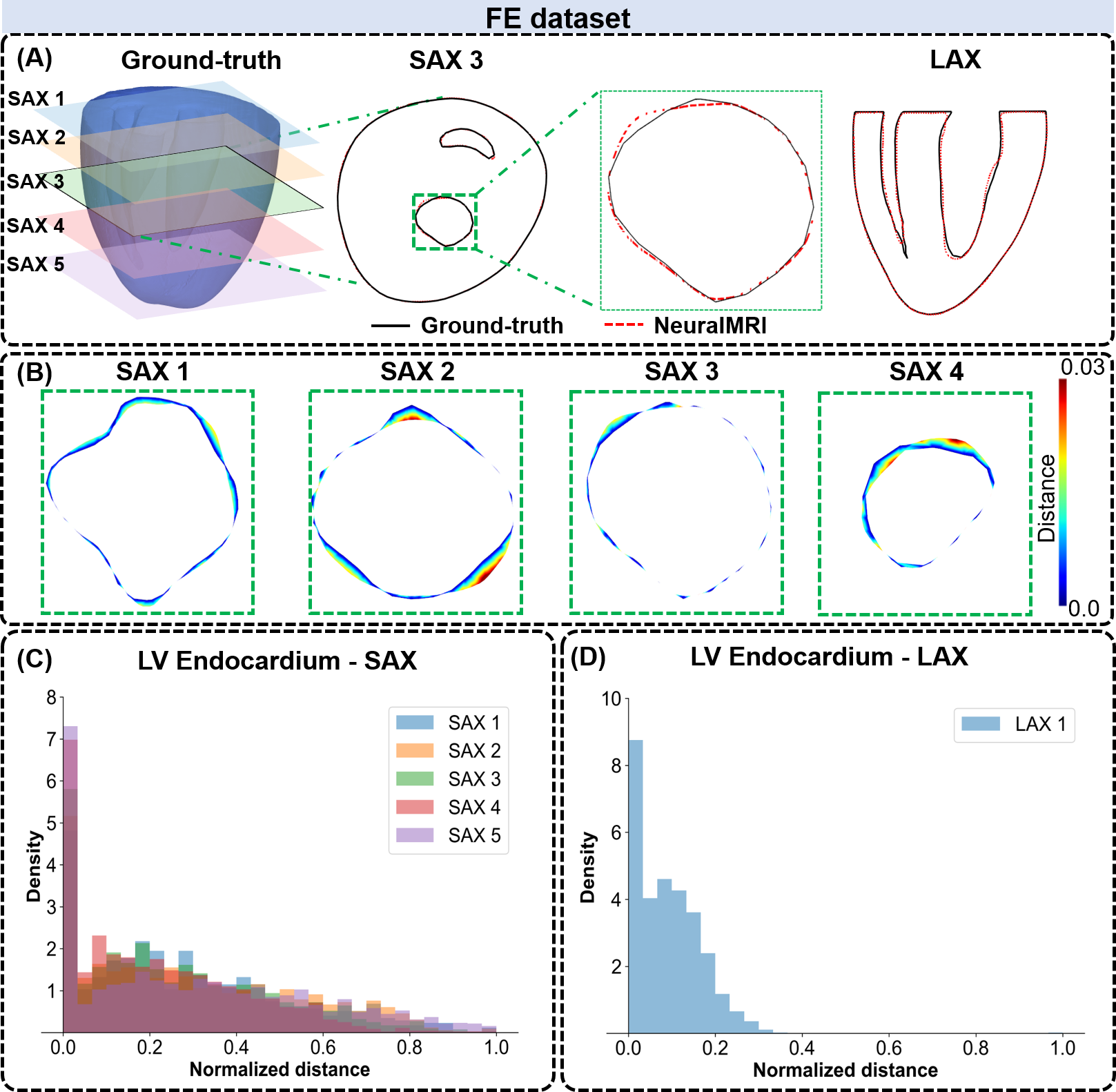}
    
    \caption{Comparison of ground-truth and reconstructed contours at multiple short- and long-axis (SAX and LAX) planes for FE dataset. (A) Location of SAX and LAX planes used for the comparison, highlighting the contours at the LV endocardium. (B) Error map of the closest distance between points in both sets of endocardial contours. (C and D) Histogram detailing the distribution of errors at different points in SAX and LAX planes, respectively. Points were normalized prior to quantitative comparison.}
    \label{fig:FE_boundary_comparison}
\end{figure}

\subsection{Comparison of distance metrics for shape reconstruction}

Quantitative and qualitative evaluation of NeuralMRI demonstrates substantial performance differences between the two sampling protocols, with dense sampling achieving sub-millimeter geometric accuracy suitable for computational simulations.
Comprehensive performance metrics for the validation split for both datasets are summarized (Table \ref{tab:table_1}). 
\noindent These metrics correspond to the highest level of sampling utilized to represent the overall neural field distribution in training the model.
NeuralMRI performed better on the single-ventricle in vivo dataset, achieving significantly higher reconstruction accuracy than on the biventricular FE dataset. The RMSE and MAE for the in vivo dataset indicate substantially lower geometric deviation than for the FE dataset (RMSE: in vivo vs. FE: 0.0412 vs. 1.420; MAE: 1.4250 vs. 0.6470, respectively). 
The Hausdorff metric captured larger vertex-wise positional deviations than the Chamfer distance, highlighting its greater sensitivity to local outliers (Table \ref{tab:table_1}). These results indicate sub-millimeter accuracy in both distance error metrics between $P^\ast$ and $P_\theta$ for the in vivo dataset.  Despite this heightened sensitivity, the 90$^{\text{th}}$ percentile window of the Hausdorff distribution was selected as the optimal threshold, as it remained tightly concentrated (only 0.5 mm above the mean).

\begin{figure}[h!]
\centering
\includegraphics[width = 0.8\linewidth]{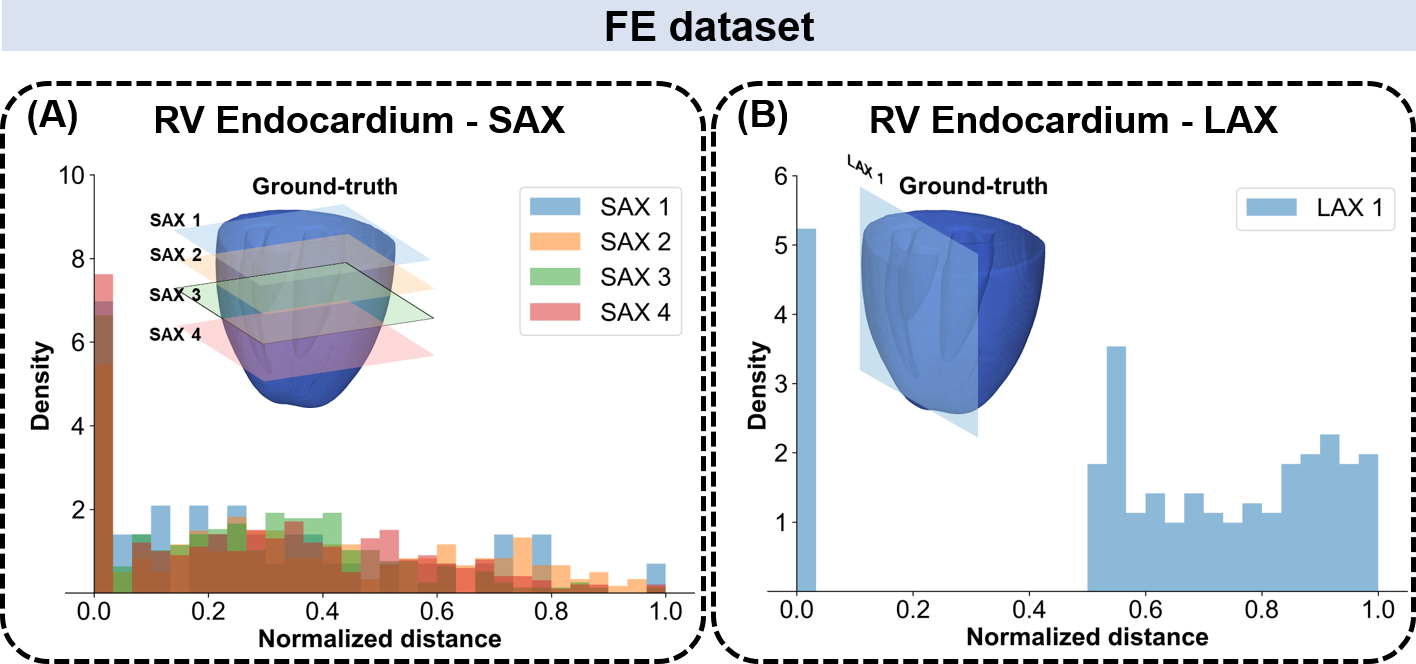}
    \setlength{\belowcaptionskip}{-16pt}
    \caption{Comparison of ground-truth and reconstructed contours of the right ventricular endocardium at multiple short- and long-axis (SAX and LAX) planes for the FE dataset. (A and B) Histogram detailing the distribution of bidirectional chamfer distance  at different points in SAX and LAX planes, respectively. Points were normalized prior to quantitative comprison. }
    \label{fig:supp_fig_1}
    
\end{figure}


\subsection{Evaluation of the geometric correspondence between ground-truth and predicted contours}

\begin{figure}[h!]
\centering
\includegraphics[width = 0.8\columnwidth]{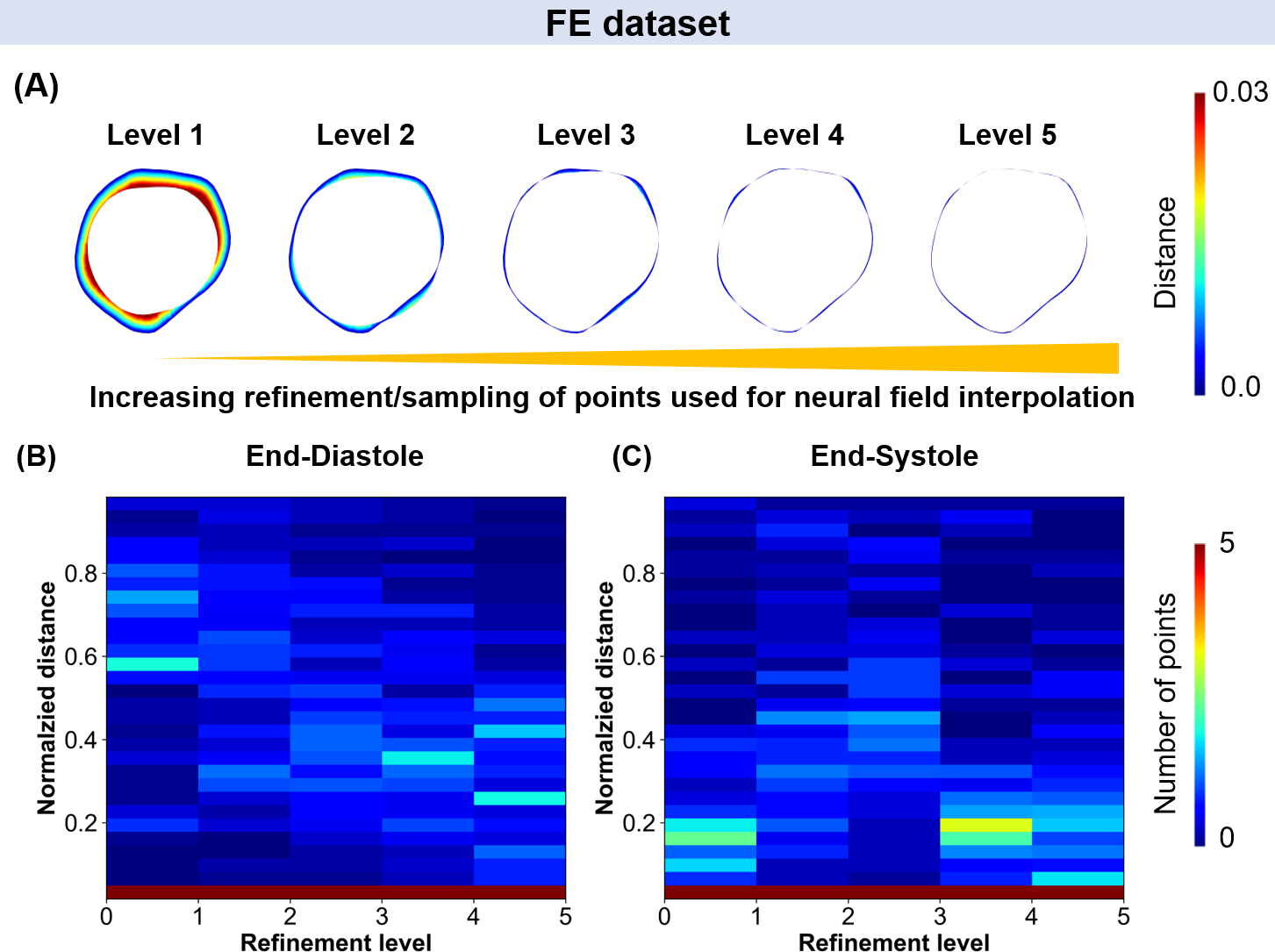}
    \setlength{\belowcaptionskip}{-10pt}
    \caption{Comparison of ground-truth and reconstructed contours of the left ventricular endocardium at different refinement levels for the FE dataset. (A) Heatmaps highlighting the bidirectional distance between the reconstrcuted LV endocardium at a specific refinement level versus the ground-truth. (B and C) Histogram of the heatmaps showing the relative changes in the distance-based error metric with each level at end-diastole and end-systole, respectively. Increasing levels correspond to a greater number of points used to represent the ground-truth geometry for neural field interpolation.}
    \label{fig:supp_fig_2}
\end{figure}

\begin{figure}[h!]
    \centering
    \includegraphics[width=0.8\columnwidth]{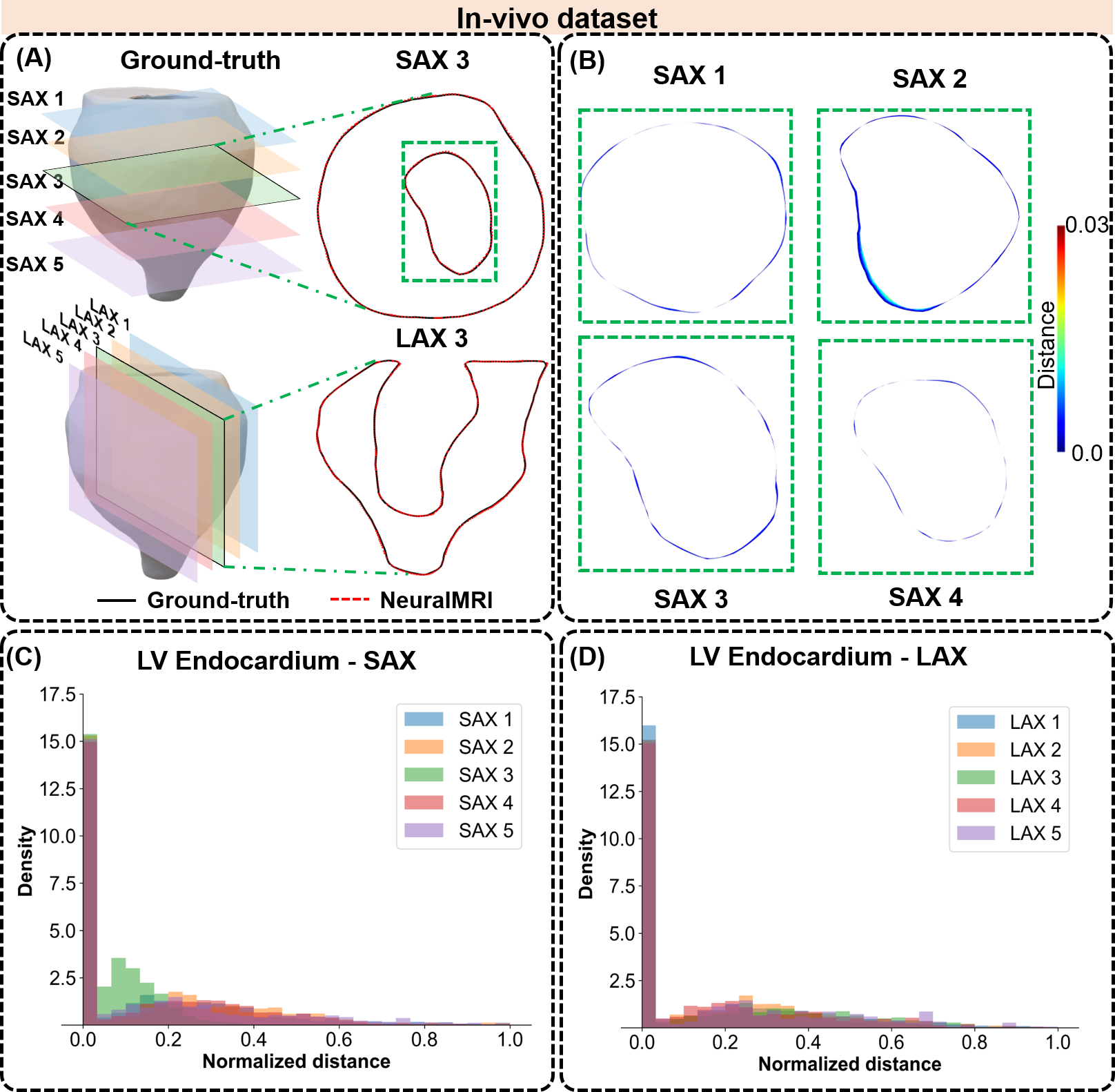}
    
    \caption{Comparison of ground-truth and reconstructed contours at multiple short- and long-axis (SAX and LAX) planes for the in vivo dataset. (A) Location of SAX and LAX planes used for the comparison, highlighting the contours at the LV endocardium. (B) Error map of the closest distance between points in both sets of endocardial contours. (C and D) Histogram detailing the distribution of errors at different points in SAX and LAX planes, respectively. Points were normalized prior to quantitative comparison.}
    \label{fig:InVivo_boundary_comparison}
\end{figure}

Qualitative assessment through planar cross-sectional comparisons at a single location of the SAX plane corroborates the overall quantitative findings of the Hausdorff and chamfer distance metrics (Figs. \ref{fig:FE_boundary_comparison}-\ref{fig:InVivo_boundary_comparison}). 
For the FE dataset, visual inspection reveals close geometric alignment in the mid-ventricular region where the density of the sampled query points is the highest in both the LV and the RV (Figs. \ref{fig:FE_boundary_comparison} and \ref{fig:supp_fig_1}). 
Increased deviations observed near the apex and base where the sampling of contour points is sparser (Fig. \ref{fig:FE_boundary_comparison}B-D).  These observations were supported by an increase in the similarity between the ground-truth and predicted contours with increasing refinement levels (Fig. \ref{fig:supp_fig_2}).
The in vivo dataset demonstrates substantially tighter boundary correspondence across all imaging planes, with ground-truth and reconstructed boundaries exhibiting minimal visible deviation, even in anatomically challenging regions (Fig. \ref{fig:InVivo_boundary_comparison}A and B). This improved alignment is consistent with across the SAX and LAX planes with the maximum errors observed in the apical region of the LV (Fig. \ref{fig:InVivo_boundary_comparison}C and D).

\subsection{Regional distribution of peak geometric correspondence errors}

The maximum deviations in geometric correspondence between $S^{\ast}$ and $S$ were statistically defined using the 90$^\text{th}$ percentile intervals of the Hausdorff distance errors, and were quantified across all points describing $S$. The in vivo dataset maintained significantly lower median error than the FE dataset throughout the entire percentile range.

\begin{figure}[h!]
    \centering
    \includegraphics[width=0.7\columnwidth]{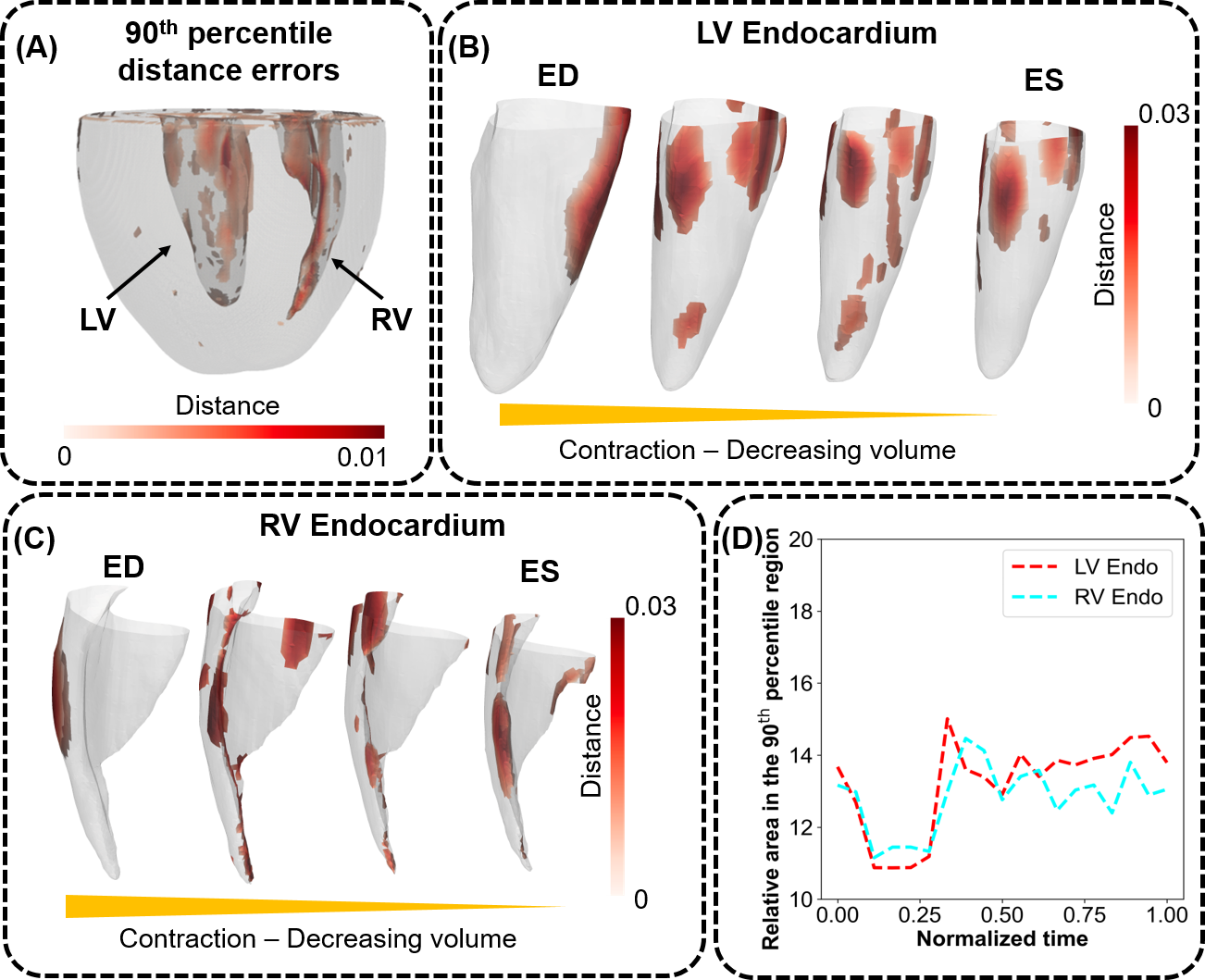}
    \caption{(A) Spatial distribution of the chamfer distance, highlighting errors within the 90$^\text{th}$ percentile region for a representative biventricular geometry from the FE dataset. (B, C) Temporal evolution of endocardial surface errors between end-diastole and end-systole for the left and right ventricles, respectively. (D) Temporal changes in the surface area of the endocardial geometries corresponding to the 90$^\text{th}$ percentile error region. The maximum error was computed based on the geometric correspondence between the ground truth and the reconstructed 3D geometries.}
    \label{fig:isosurface_distance_comparison}
\end{figure}

\noindent Approximately 90\% of surface points in in vivo dataset exhibited reconstruction errors below 0.1 mm, confirming clinically acceptable accuracy in generating the ground-truth LV. 
\noindent As such, results are visualized for the FE datasets, with iso-surfaces constructed from the points defining the 90$^\text{th}$ percentile intervals (Fig. \ref{fig:isosurface_distance_comparison}). 
The reconstruction of the biventricular geometries exhibited broader error distributions and greater variability, particularly near the LV and RV chambers. Whereas the errors in the LV chamber were concentrated towards the base and apex, the spread was broader in the RV (Fig. \ref{fig:isosurface_distance_comparison}B and C).  
Maximum errors were observed during peak systolic contraction, with the total range between 12$\%$ and 18$\%$ of the relative area forming the LV and RV shapes (Fig. \ref{fig:isosurface_distance_comparison}D). 
The temporal changes in error-prone regions were evaluated using a histogram for both datasets (Fig. \ref{fig:cummulative_distance_percentile_comparison}). The error distribution characteristics directly correlate with sampling density, with the in vivo dataset's comprehensive multi-plane acquisition providing sufficient geometric constraints to accurately capture complex anatomical features that remain underresolved in the sparse sampling protocol, i.e., the non-convex contours of the RV (Figs. \ref{fig:cummulative_distance_percentile_comparison}A and B). 
Despite the reduction in the perceived concavity of the RV isosurface with diastolic relaxation (ES to ED), the relative area covered by the 90$^\text{th}$ percentile intervals remained unperturbed. These results reflected challenges in reconstructing regions with complex geometry or insufficient sampling density.

\subsection{Time-dependent changes in curvature error distribution}
\begin{figure}[h!]
\centering
\includegraphics[width = 0.75\linewidth]{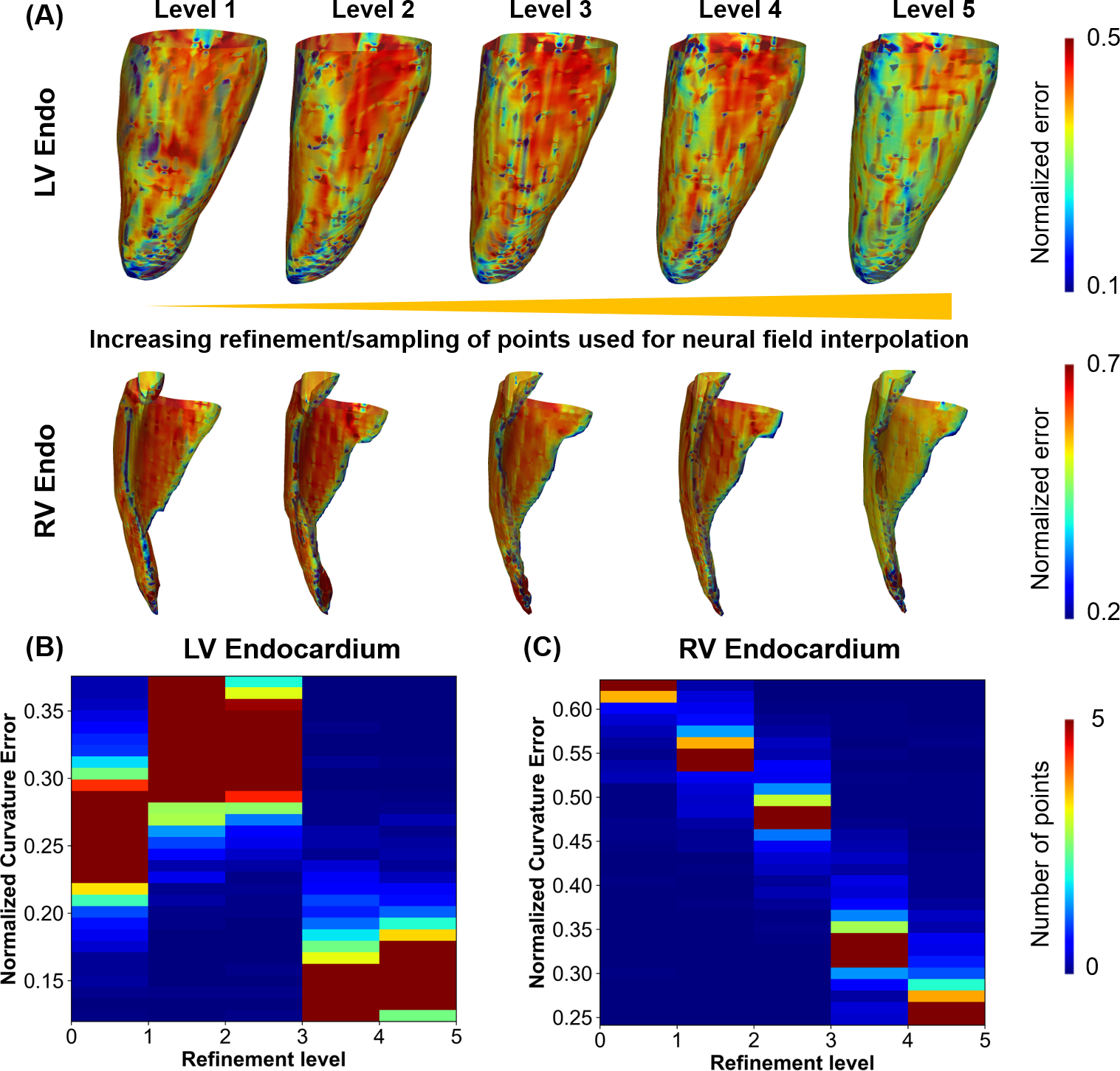}
    \setlength{\belowcaptionskip}{-16pt}
    \caption{(A) Comparison of ground-truth and reconstructed error in the curvature metric between the ground-truth and reconstructed 3D geometries in the FE dataset at different refinement levels. (B and C) Histogram of the heatmaps showing the relative changes in the curvature error metric metric with each level at the LV and RV endocardium. Increasing levels correspond to a greater number of points used to represent the ground-truth geometry for neural field interpolation.}
    \label{fig:supp_fig_3}
    
\end{figure}

 The comprehensive multi-plane contour positioning provided sufficient geometric constraints to preserve local curvature features. The regional distribution of the local curvature of $q \in P_{\theta}$ was qualitatively similar to $v \in V^{\ast}$ (Fig. \ref{fig:curvature_distribution}). Results are visualized as heatmaps for the LV and RV endocardial surfaces (Fig. \ref{fig:curv_error_maps}A and B). Despite these similarities, the sparse sampling protocol failed to fully resolve regions of high curvature, particularly along the non-convex contours of the RV even at the highest refinement level (Fig. \ref{fig:supp_fig_3}). Although diastolic relaxation (ES to ED) reduced the apparent concavity of the reconstructed RV isosurface, the relative surface area encompassed by the 75$^{\text{th}}$ percentile curvature error intervals remained largely unchanged (Figs. \ref{fig:curv_error_maps}C and D). 

\section{Discussion}
\label{sec:Discussion}

\subsection{Global implicit representations improve shape reconstruction under sparse sampling}
This study demonstrates that a neural field–based reconstruction pipeline can accurately recover 3D heart geometries from sparse planar contour data while preserving regional anatomical detail. Reconstruction performance was evaluated on two datasets, comprising in-silico and in vivo simulations that varied sampling density and geometric complexity. In both cases, the reconstructed meshes closely matched reference geometries, indicating that the learned implicit representation faithfully captures cardiac planar contours. Unlike traditional spline-based interpolation, which fits contours locally and is prone to artifacts in regions of high curvature or irregular sampling, the neural field approach learns a global, continuous representation of shape \cite{Wang-2025, Dyrby-2014}. By leveraging spatial context across all input planes, this framework enables robust interpolation in anatomically challenging regions, such as the apex and basal sections of the LV endocardium, thereby improving geometric fidelity under sparse sampling conditions. Error analysis revealed that while average geometric deviations remained low for densely sampled data, local outliers persisted in anatomically challenging regions (Fig. \ref{fig:FE_boundary_comparison}). Metrics sensitive to vertex-wise deviations, such as the Hausdorff distance, highlighted these localized errors more strongly than global measures, motivating the use of percentile-based thresholds to characterize maximal reconstruction uncertainty. Spatial and temporal analyses further showed that reconstruction errors clustered near the LV base, apex, and RV free wall—regions known to exhibit high curvature and non-convex morphology. Collectively, these findings indicate that neural field reconstructions are robust under dense sampling but remain sensitive to undersampling in anatomically complex regions, underscoring the importance of sampling density as a primary determinant of reconstruction fidelity.

\subsection{Geometry-specific training improved through-plane reconstruction of complex shape features}
The subject-specific approach enhanced geometric correspondence of planar contours, leading to improved iso-surface reconstruction of the LV endocardium in both datasets. A standard limitation of MLPs is the appearance of smoothing artifacts that may manifest as geometric dilation of an isosurface. This is especially true in the context of 3D point positional encoding or geometric occupancy. 
\begin{figure}[h!]
    \centering
    \includegraphics[width=0.8\columnwidth]{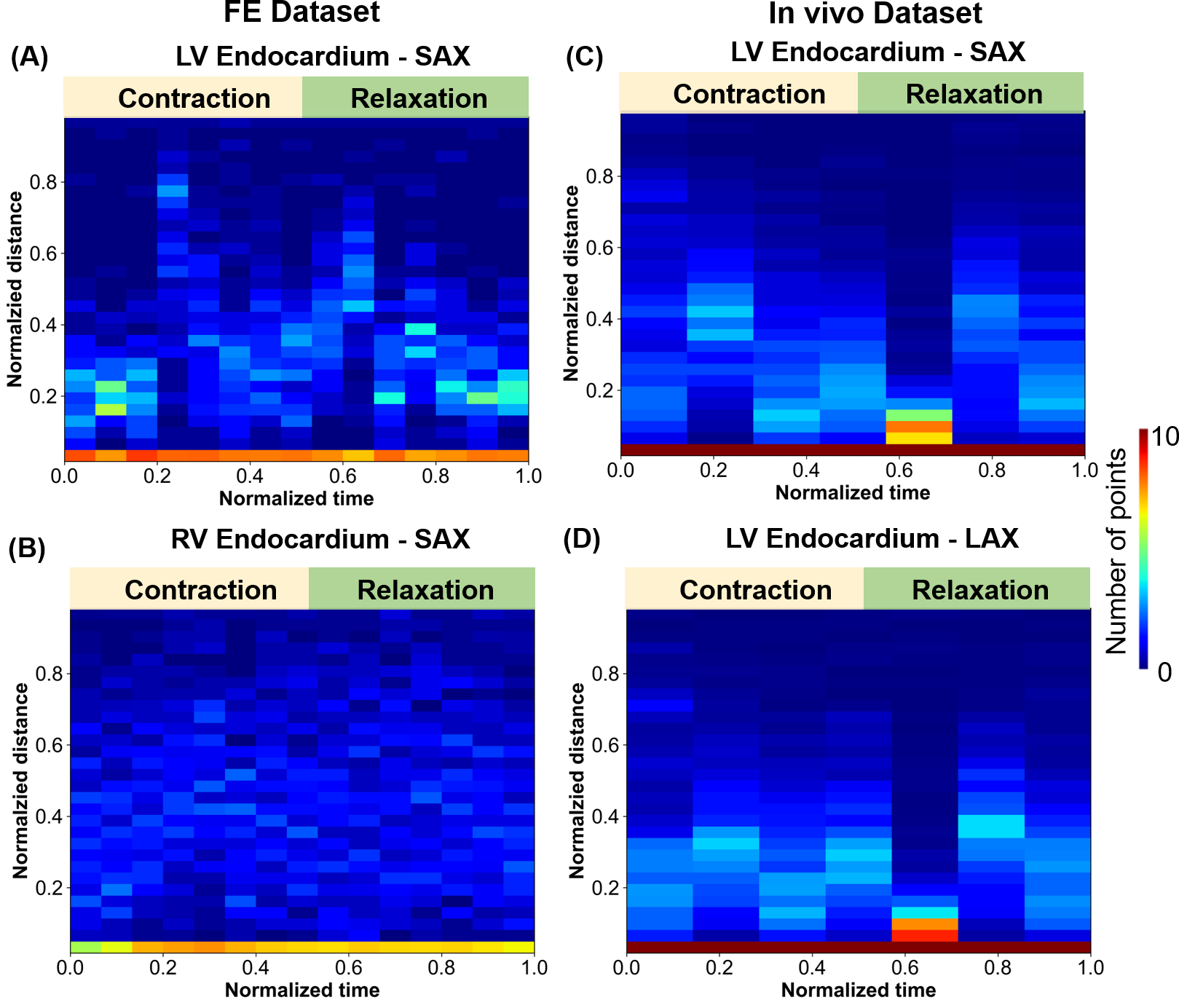}
    \setlength{\belowcaptionskip}{6pt}
    \caption{ Histograms detailing the temporal evolution and spatial distribution of the chamfer distance, highlighting errors within the 90$^\text{th}$ percentile region for all geometries in the (A and B) FE or synthetic and (C and D) In vivo datasets. Errors are presented as changes to the surface area of the endocardial geometries corresponding to the 90$^\text{th}$ percentile error region. The maximum error was computed based on the geometric correspondence between the ground truth and the reconstructed 3D geometries.}
    \label{fig:cummulative_distance_percentile_comparison}
\end{figure}

\begin{figure}[h!]
    \centering
    \includegraphics[width=0.8\columnwidth]{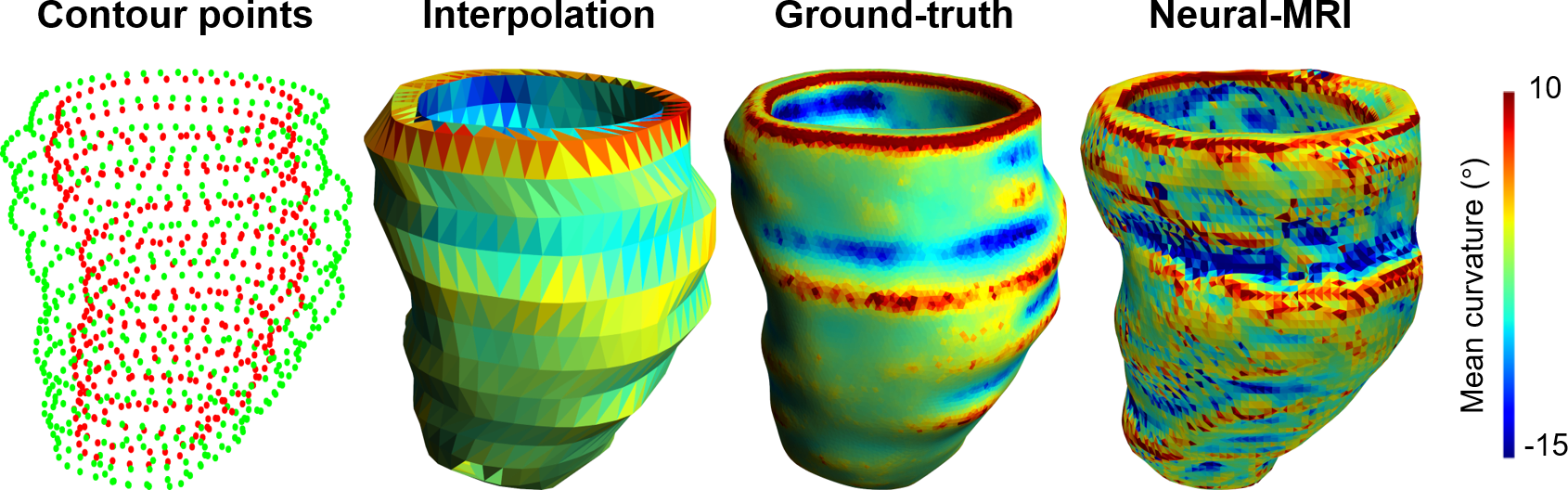}
    \caption{Spatial distribution of the mean curvature metric for a representative human-specific left ventricle geometry from the FE dataset. (left to right) Initial contours of the endo - and epicardium segmented using short-axis imaging, followed by reconstruction through (i) linear interpolation, (ii) spline-based interpolation and adaptive remeshing  (ground-truth), and (iii) Neural-MRI}
    \label{fig:curvature_distribution}
\end{figure}
The input data used in NeuralMRI comprises a vast network of arbitrarily aligned image-contour pairs with minimal areas of intersection to represent a 3D space with the iso-surface comprising multiple geometric features, including curves, folds, and edges (Figs. \ref{fig:FE_boundary_comparison} and \ref{fig:InVivo_boundary_comparison}). Indeed, encoding the topology of the biventricular heart often requires a substantial training dataset of multiview images to capture the full 3D space. The NeuralMRI model was first trained directly on positional and shape features using iso-surfaces obtained from rigorous in-silico simulations of a heart model phantom, and subsequently back-propagated the learned sampling-point representation into the original latent space, thereby improving training fidelity. Subsequently, the decoder of the MLP architecture integrates sampling occupancy representation (shape features) with the encoded contour pairs (positional features.) The neural field interpolator model improved correspondence with increasing levels of refinement, thus indicating a closer approximation of the ground-truth contours, even in cases not included during training (Fig. \ref{fig:cummulative_distance_percentile_comparison}). Despite increasing stability, unrealistic geometric features---including dilation and folds---could have resulted in shapes that are not representative of the local curvature of the ground-truth RV shape. Recently, multi-fidelity training has been identified as the potential solution for generating realistic features \cite{Renzi-2025, Sajjadinia-2022}. Subsequent efforts will combine labeled cardiac MRI data to guide the generation of 3D heart models that are more consistent with expected anatomical structures \cite{Catalan-2025}. Nevertheless, the proposed approach may overcome limitations in generating accurate LV shape representations for various computational tasks. 

\subsection{Multi-view segmentation to facilitate clinical translation}

Multi-view contour sampling significantly enhances the quality of geometric reconstruction by combining clinically relevant anatomical information from SAX and LAX imaging planes. The neural field architecture integrates arbitrarily oriented cross-sections into a unified spatial representation, whereas the positional encoding approach associates spatial coordinates with clinically feasible scanning planes, enabling rapid cardiac shape reconstruction. The superior performance of the in vivo dataset can be attributed in part to more comprehensive multi-view sampling compared to the FE dataset (Fig. \ref{fig:cummulative_distance_percentile_comparison}).
\begin{figure}[h!]
    \centering
    \includegraphics[width=0.8\columnwidth]{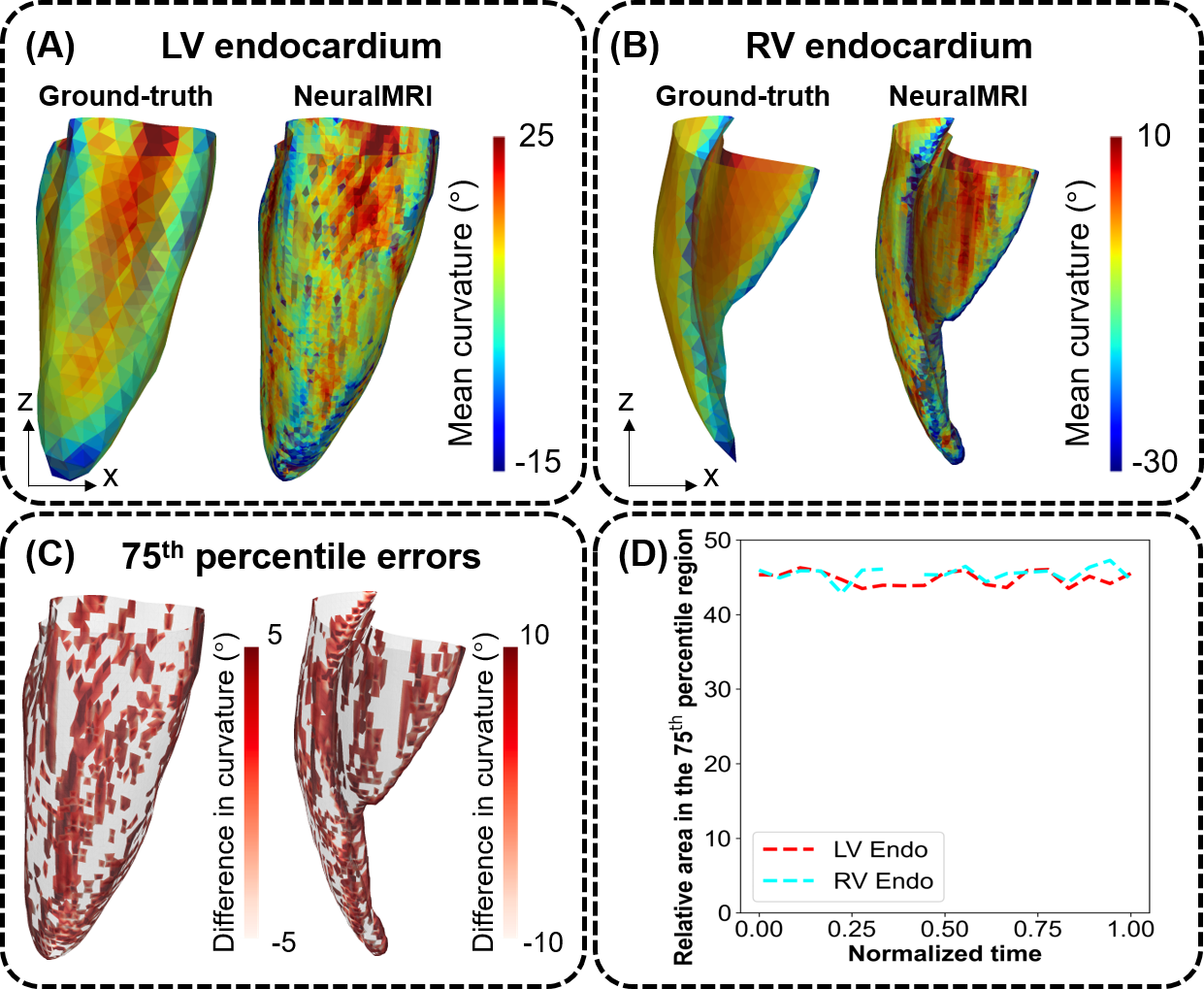}
    \caption{Spatial distribution of the error in the curvature metric between the ground-truth and reconstructed 3D geometries at the (A) LV and (B) RV endocardium, respectively, from the FE dataset. (C) Error distribution within the 75$^\text{th}$ percentile region for a representative biventricular geometry. (D) Temporal changes in the surface area of the endocardial geometries corresponding to the 75$^\text{th}$ percentile error region. Maximum error was computed based on geometric correspondence.}
    \label{fig:curv_error_maps}
\end{figure}
The cardiac shape was subsequently reconstructed as an isosurface using the marching cubes algorithm and was not included in the training process. Despite challenges such as unrealistic geometric distortions and the limitations of viewpoint-based encoding, the proposed approach represents a promising step toward achieving more accurate and anatomically consistent LV iso-surfaces.  Our method uses a viewpoint-encoding approach to learn an implicit field representation, i.e., by associating hierarchical sampling with contour positions. Indeed, neural implicit fields are known to provide numerically stable representations of sharp geometric features \cite{Amiranashvili-2024}, such as the local curvatures of the in vivo LV in this study (Fig. \ref{fig:curvature_distribution}). 
Despite qualitative similarity between the ground-truth and predicted geometries, curvature errors were encountered while validating the reconstruction of the RV endocardium (Fig. \ref{fig:curv_error_maps}). Our subsequent efforts will include time-wise image encoding using recurrent networks or transformers to learn temporal features and improve the realism of geometric heart features \cite{Lyu-2023, Vornehm-2025}. As such, learning iso-surface reconstruction often involves learning geometric features of the heart from a sequence of LR images to manipulate local shape features. The multi-plane integration strategy may be particularly effective for capturing out-of-plane motion and geometric features that are poorly resolved from single-view acquisitions alone. By incorporating point-set registration of cardiac motion-derived geometries, we expect the neural fields to capture out-of-plane motion adjustments, thereby providing a physically detailed approach to 3D cardiac reconstruction, as seen in other studies \cite{Sun-2024}.



\subsection{Limitations and Future Directions}

While the proposed method demonstrates promising reconstruction capabilities, a few limitations warrant discussion. First, the models in this study were restricted to a combined in vivo dataset of ventricular geometries in mouse and human hearts. However, accurate digital twins for electromechanical and electrophysiological simulations require anatomically precise representations that vary by subject. Most segmented images used here were derived from motion-mapping studies, which primarily focus on generating the closest approximation of the convex LV shape. Future work will incorporate individual datasets of murine and human-specific geometries to improve the physiological fidelity of neuralMRI. Second, the current framework does not explicitly enforce temporal coherence across the cardiac cycle. Incorporating recurrent neural networks or transformer-based architectures could enable the model to learn dynamic cardiac motion patterns and improve temporal consistency across phases. Although FE simulations were used to model contractile kinematics, these datasets were used independently from in vivo data for training, testing, and validation. This separation contributed to disparities in error distributions between single-ventricle and biventricular reconstructions. 
While diastolic relaxation reduced apparent geometric complexity, high-error regions persisted in the FE simulation-derived biventricular dataset, suggesting that reconstruction inaccuracies were driven primarily by insufficient geometric constraints rather than cardiac phase. Future work will incorporate multi-fidelity training strategies by integrating labeled cardiac MRI datasets with physics-based simulations, thereby improving the generation of anatomically accurate 3D heart models and reducing geometric distortions in challenging anatomical regions.

\section{Data availability} \label{sec:Data}
Code and supporting data, including animations and results, are available in the 'NeuralMRI' Github repository (https://github.com/Tanmay24Mukh/NeuralMRI.git). Any additional data that support the findings of this study are available from the corresponding author, R.A., upon request.

\section{Acknowledgements} \label{sec:Acknowledgements}
This work was supported by National Institutes of Health R01HL168368 to R.A and American Heart Association 24PRE1240097 to T.M.


\end{document}